\documentclass[pdflatex,sn-mathphys-num,iicol]{sn-jnl}

\emergencystretch=\maxdimen
\usepackage{graphicx}
\usepackage{amsmath}
\usepackage{url}
\usepackage{float}
\usepackage{xcolor}
\usepackage{accessibility}
\usepackage{comment}   
\usepackage{algorithm,algpseudocode}
\usepackage{multirow}
\usepackage{tabularx}
\usepackage{xcolor}
\usepackage{rotating}

\usepackage{subcaption}
\newdimen{\algindent}
\newcommand{\commentsymbol}{//}
\algrenewcommand\algorithmiccomment[1]{\hfill \commentsymbol{} #1}
\newdimen{\algindent}
\usepackage{varwidth}
\usepackage{xcolor}
\usepackage{mdframed}
\usepackage{booktabs}
\usepackage{wrapfig}
\usepackage{float}

\usepackage{amssymb}  
\usepackage{balance}

\mdfdefinestyle{MyFrame}{
    linecolor=black,
    outerlinewidth=2pt,
    innertopmargin=4pt,
    innerbottommargin=4pt,
    innerrightmargin=4pt,
    innerleftmargin=4pt,
    leftmargin = 4pt,
    rightmargin = 4pt,
    backgroundcolor=white
}
\newcounter{insight}
\newenvironment{insight}{
    \stepcounter{insight}
    \vspace{3mm}
    \begin{mdframed}[style=MyFrame]\itshape
    \textbf{Finding~\theinsight:}
    }
    {
    \end{mdframed}
    \vspace{10pt}
    }

\newcommand{\toolName}{Senior Fit}

\begin{document}

\title{What Does It Take? Developing a Smartphone App that Motivates Older Adults to be Physically Active}

\author*[1]{Sabrina Haque}\email{sxh3912@mavs.uta.edu}
\author[1]{Kyle Henry}\email{kyle.henry@mavs.uta.edu}
\author[2]{Troyee Saha}\email{txs2850@mavs.uta.edu}
\author[3]{Kimberly Vanhoose}\email{kimberly.vanhoose@mavs.uta.edu}
\author[2]{Jobaidul Boni}\email{jobaidulalam.boni@mavs.uta.edu}
\author[3]{Samantha Moss}\email{samantha.moss@unt.edu} 
\author[2]{Kate Hyun}\email{kate.hyun@uta.edu}
\author[4]{Kathy Siepker}\email{kathy.siepker@uta.edu}
\author[3]{Xiangli Gu}\email{xiangli.gu@uta.edu}
\author[5]{Angela Liegey-Dougall}\email{adougall@uta.edu}
\author[2]{Stephen Mattingly}\email{mattingly@uta.edu}
\author[1]{Christoph Csallner}\email{csallner@uta.edu}

\affil[1]{\orgdiv{Department of Computer Science and Engineering}, \orgname{University of Texas at Arlington}, \orgaddress{\city{Arlington}, \postcode{76019}, \state{Texas}, \country{USA}}}

\affil[2]{\orgdiv{Department of Civil Engineering}, \orgname{University of Texas at Arlington}, \orgaddress{\city{Arlington}, \postcode{76019}, \state{Texas}, \country{USA}}}

\affil[3]{\orgdiv{Department of Kinesiology}, \orgname{University of Texas at Arlington}, \orgaddress{\city{Arlington}, \postcode{76019}, \state{Texas}, \country{USA}}}

\affil[4]{\orgdiv{School of Social Work}, \orgname{University of Texas at Arlington}, \orgaddress{\city{Arlington}, \postcode{76019}, \state{Texas}, \country{USA}}}

\affil[5]{\orgdiv{Department of Psychology}, \orgname{University of Texas at Arlington}, \orgaddress{\city{Arlington}, \postcode{76019}, \state{Texas}, \country{USA}}}

\abstract{
Maintaining physical activity is essential for older adults' health and well-being, yet participation remains low. Traditional paper-based and in-person interventions have been effective but face scalability issues. Smartphone apps offer a potential solution, but their effectiveness in real-world use remains underexplored. Most prior studies take place in controlled environments, use specialized hardware, or rely on in-person training sessions or researcher-led setup. 
This study examines the feasibility and engagement of \toolName{}, a standalone mobile fitness app designed for older adults. We conducted continuous testing with 25 participants aged 65--85, refining the app based on their feedback to improve usability and accessibility. 
Our findings underscore both the potential and key challenges in designing digital health interventions. Older adults valued features such as video demonstrations and reminders that made activity feel accessible and motivating, yet some expressed frustration with manual logging and limited personalization. The Facebook group provided encouragement for some but excluded others unfamiliar with the platform. These results highlight the need for fitness apps that integrate flexible tracking, clear feedback, and low-barrier social support. We contribute design recommendations for creating inclusive mobile fitness tools that align with older adults’ routines and capabilities, offering insights for future long-term, real-world deployments.}


\maketitle

\section{Introduction}

The global population is growing rapidly, with over 2 billion people expected to be aged 60 or older by 2050~\cite{wheatley2005discipline}, and 80\% of whom will be residing in low- and middle-income countries~\cite{who_2022}. In the United States, approximately 27\% of adults aged 60 and older live alone, a figure that rises to 43\% for women of ages~75 or above~\cite{census}. While many older adults prefer to age in place, i.e., live in their own home and community safely, independently, and comfortably~\cite{aarp}, doing so requires maintaining mobility, independence, and overall well-being.

Aging populations face increasing risks of physical inactivity and social isolation, which both contribute to declining well-being~\cite{wheatley2005discipline,gard2016world}. Physical activity is an important component of an active and healthy lifestyle. Regular exercise slows age-related health decline, improves mobility, cognitive function, and overall well-being~\cite{jones2010natural,baltes1993successful,chodzko2009exercise}. Conversely, sedentary behavior is a major contributor to chronic diseases worldwide~\cite{ozemek2019global}. Despite these benefits, many older adults struggle to stay active, often due to a lack of motivation, physical limitations, or different accessibility barriers~\cite{nelson2007physical,baert2011motivators}. The COVID-19 pandemic further increased sedentary behavior across populations~\cite{hall2020tale} and emphasized the urgency of developing fitness solutions that are remote-friendly, sustainable, and inclusive~\cite{who_covid,banerjee2020estimating}.

Over the past several decades, many efforts have been taken to promote physical activity among older adults. Early approaches mostly took the form of in-person, group-based exercise programs, often conducted in clinical or community settings~\cite{king1998physical,short2011efficacy}. These programs provided structured routines, personalized feedback via face-to-face interactions with professionals, and valuable social support~\cite{taylor2004physical,daskalopoulou2017physical,richards2013face}. However, they suffered from limited scalability and often excluded individuals with transportation or scheduling constraints~\cite{king1998physical}. In-person physiotherapy, although effective, often demands travel, scheduling commitments, and financial costs, making them inaccessible for individuals with mobility issues or limited resources~\cite{argent2018patient}. Home-based programs using mailed materials and periodic check-ins reduced these barriers, but faced challenges in sustaining long-term engagement. Participants often fall back into their older habits once the structured program ended~\cite{van2002effectiveness}. Besides, paper-based programs also introduced practical limitations, requiring participants to manage physical reports, and mailing delays created additional burdens and administrative costs~\cite{marcus2007comparison,short2011efficacy}. The COVID-19 pandemic further disrupted access to community-based exercise programs, leaving many older adults without structured support for maintaining physical activity~\cite{krendl2021impact,kasar2021life}.

Digital health interventions have risen as a promising alternative. Many rely on specialized hardware such as pedometers, fitness trackers, smartwatches, or even augmented/virtual reality systems~\cite{chen2024silvercycling,androutsou2020smartphone}. These tools can deliver passive tracking or interactive exercise experiences, but they often introduce barriers of their own: cost, technical setup, frequent charging, and assumptions about digital literacy~\cite{moore2021older}. In many studies, such devices are typically mailed to participants with pre-configured settings and instructions, which can ease initial use challenges, but does not reflect the reality of older adults adopting these tools on their own ~\cite{mair2022personalized}. Moreover, many of the systems may also frame older adults primarily as patients to be monitored, for example, for fall detection or adherence, rather than as active participants in their own health, which can undermine motivation and enjoyment~\cite{vargemidis2020wearable,vargemidis2021irrelevant}.

A growing body of work has explored smartphone or tablet apps without requiring additional external hardware or immersive technologies. These apps typically deliver structured exercise routines, instructional videos, or behavior change features such as goal setting, reminders, and motivational messaging~\cite{kaur2022usability,paul2017increasing,aslam2020systematic}, sometimes supplemented with phone calls, text messages, or video-conferencing to sustain motivation and reinforce healthy behaviors~\cite{li2022cognitively,daly2021feasibility}. Compared to in-person or equipment-intensive interventions, mobile apps offer a more lightweight and potentially scalable path for promoting physical activity at home, utilizing the increasing smartphone adoption among older adults~\cite{techuse,smartphoneuse}. The effectiveness of these apps in real-world, long-term use, however, remains unclear. 
Most prior studies have taken place in controlled environments or short-term field deployments, often with direct researcher involvement~\cite{mehra2019supporting,papi2020feasibility}. In many cases, researchers provide in-person onboarding, install the app on participants’ devices, and offer guidance for completing complex tasks. These conditions provide limited insight into whether older adults can independently sustain app use over months. Furthermore, many such studies are designed to support very specific physical goals, such as fall prevention,  balance improvement, or walking~\cite{daniels2025exploring,takahashi2016mobile,mostajeran2020augmented}, which may limit broader utility for older adults who are trying to stay generally active, without any specific diagnosis or supervised goal. Several also adopt a deficit-oriented view, treating older adults as passive recipients of care rather than as active participants in their own health journeys~\cite{gerling2020critical}. 

Usability challenges further limit the adoption of this approach. Even though smartphone use among older adults is steadily increasing, many older adults still find existing fitness apps difficult to use, or hard to stick with over time~\cite{martin2023importance}. This is often because the apps feel irrelevant, intrusive, or just plain overwhelming~\cite{baer2022middle}. Common design flaws, such as tiny tap targets, small fonts, cluttered layouts, and gestures like drag-and-drop, can make navigation frustrating~\cite{kaur2022usability,franklin2018engagement}. Many commercial fitness and cognitive training apps often fail to account for age-related differences in vision, dexterity, and cognitive load. In one evaluation of Nike+ and Runkeeper, both popular fitness apps aimed at promoting physical activity, researchers found that they violated basic design principles for older adults, including insufficient contrast, reduced font size, and poor touch accessibility~\cite{silva2014smartphones}. These issues are especially tough for users without much prior experience using digital tools, requiring additional support~\cite{quinn2019mobile,kruse2017mobile}. Recent work has therefore emphasized participatory and co-design approaches, which involve older adults directly in shaping technologies to reflect their everyday practices, values, and needs~\cite{harrington2018designing, so2024they}. 

To explore the feasibility of a scalable, technology-driven solution to support older adults in staying active while aging at home, we developed \toolName{}, a lightweight, smartphone-based fitness app. \toolName{} is designed to be a standalone app, without requiring wearables or external hardware, so that \textbf{participants can install and use \toolName{} independently on their own smartphones}. Unlike interventions targeting a single health issue, \toolName{} offers a comprehensive set of features---indoor exercises with video demonstrations and automatic exercise tracking, walk tracking, motivational notifications, and a private Facebook group for peer support. 

We conducted a six-month, unsupervised field study where 25 older adults used the \toolName{} app in their everyday lives in a self-guided manner. While we provided support through text, phone, or email when needed, \textbf{we did not conduct any in-person training or onboarding}, to evaluate how the app performed under realistic conditions. We follow a mixed-methods evaluation by combining app usage logs, participant surveys, and interviews. Through this long-term, real-world deployment, we try to provide insights into what works, what does not, and what remains challenging when designing inclusive fitness apps for older adults.

In this context we explore the following research questions.

\begin{itemize}
\item[RQ1] How do participants engage with the \toolName{} app during the six-month deployment?
\item[RQ2] How does the app-based intervention phase compare to our earlier paper-based intervention phase in terms of supporting participants' physical activity and well-being?
\item[RQ3] What preferences and challenges do participants express regarding \toolName{}'s physical activity tracking?
\item[RQ4] To what extent do \toolName{}'s Facebook-based social support features promote interaction and engagement among participants?
\item[RQ5] How do participants adapt \toolName{} to their personal routines?
\item[RQ6] How do participants evaluate \toolName{}'s privacy practices?
\end{itemize}

\section{Methodology}

We conducted this study in two phases. In the first phase, we used more traditional methods: printed exercise plans, SMS reminders, and simple activity logs. This helped us understand what older adults respond well to and what kinds of motivation or structure they prefer, without introducing new technology right away. In the second phase, building on what we learned, we created \toolName{}, a smartphone app designed for older adults. It aimed to offer the same kind of guidance and support, but in a digital format. We wanted to see if the app could be just as effective, and possibly more flexible or scalable, especially given the growing use of smartphones and the push for remote-friendly options during the COVID-19 pandemic.

In this paper, we focus on phase~2 of our study, i.e., the smartphone app-based intervention. To do that, we briefly summarize the phase~1 aspects that are relevant for phase~2. A detailed description of phase~1 activities and results is beyond our scope and will appear in a venue focused on health and aging.

\subsection{Ethical Considerations}

This study was conducted in accordance with ethical guidelines for research involving human subjects. All research personnel completed human subjects protection training, and the study protocol was approved by our university’s Institutional Review Board (IRB). Participants provided informed consent before enrollment, ensuring their voluntary participation and a clear understanding of the study’s purpose and procedures.

\subsection{Phase-1: SMS and Paper-Based Intervention}

We recruited older adults (within the DFW metropolitan area) via a study website linking to an application form~\cite{SeniorFit_Materials}, mail outreach, flyers distributed at retirement communities, and in-person at senior events. COVID-19 pandemic restrictions complicated the recruitment of seniors (especially in-person), yielding fewer participants than planned.

Eligible participants were 64 and older English-speaking smartphone users who have used some phone apps. Incentives included Walmart gift cards (up to \$200 over nine months). We excluded participants residing in long-term care facilities to ensure the study focused on independently living older adults, who are more likely to manage their own physical activity routines and technology use. Of 140 eligible individuals, 97 completed the baseline survey and proceeded to phase~1.

\subsubsection{Baseline Survey: Tech Familiarity of 97 Phase~1 Participants}

Before initiating the phase~1 intervention, we conducted a broad six-section baseline survey~\cite{SeniorFit_Materials}, consisting of basic information, demographics, quality of life, gait and balance, physical activity, and familiarity with technology. All recruits had the option to complete this survey by mail, phone, or in-person interview. 97 phase~1 recruits (ages 64--88, avg.~72, median~71) completed this survey.

In this section, we focus on the sixth part of this survey, i.e., seniors’ familiarity with technology, especially smartphones. Our goal was to identify tech-savvy participants to test beta versions of the \toolName{} app for an additional \$20 Walmart gift card.

\begin{table*}[h!t]
\centering
\caption{Phone-based channels (``Within the last 12~months, how often have you used ... on your smartphone?''): Share of phase~1 \& 2 participants indicating 
``several times a day'' (d+), 
``about once a day/week/month'' (d, w, m), 
``never'' (0), 
or blank.
}
\begin{tabular}{l|rrrrrr|rrrrrr}
\toprule
   & \multicolumn{6}{c|}{Phase 1 (n=97)} & \multicolumn{6}{c}{Phase 2 (n=25)}\\
 (\%) & d+ & d & w & m & 0 & - & d+ & d & w & m & 0 & - \\ 
\midrule 
 Facebook & 38 & 19 & 8 & 7 & 20 & 8 & 46 & 13 & 8 & 8 & 25 & 0 \\
 YouTube & 13 & 22 & 29 & 15 & 13 & 8 & 8 & 29 & 46 & 8 & 8 & 0 \\
 Video call & 4 & 5 & 32 & 31 & 20 & 8 & 4 & 8 & 42 & 29 & 17 & 0 \\ 
 Email & 63 & 16 & 7 & 3 & 6 & 5 & 75 & 4 & 8 & 0 & 13 & 0 \\
 Text message & 76 & 12 & 4 & 2 & 1 & 5 & 88 & 0 & 4 & 4 & 4 & 0 \\
\bottomrule
\end{tabular}
\label{tab:digital_engagement_summary}
\end{table*}

In the survey, most (86/97) participants indicate they use a smartphone, which is a key requirement for our phase~2 intervention. We also polled for familiarity with a few specific apps and app categories we contemplated using in our study. Specifically, over three quarters of phase~1 participants stated that in the year before the study they used their smartphone several times a day (Table~\ref{tab:digital_engagement_summary}).
The \textbf{most widely used phone-based communication channel is SMS text messages} followed by email. Specifically, 92\% of participants report at least weekly use of SMS, followed by 86\% for email, vs. 65\% for Facebook, 64\% for YouTube, and 41\% for video calls.

We also asked about other apps (``Which other apps do you typically use at least once a week on your smartphone? (E.g.: Weather, What's app, etc.)''). The most frequently used app was Weather (43/97 participants), followed by Calendar (12), Amazon (12), Games (11), News (10), Banking (9), Google (9), Instagram (9), and Camera (9).

Several participants indicated to be willing to at least discuss installing our beta versions (``How comfortable would you be with installing on your smartphone various `under development' versions of our application? Please note that you may need to install several newer versions on your phone since it may be only partially complete and change significantly from one version to the next?''). Specifically, 7\% selected ``not comfortable'', 16\% ``I am not familiar with ‘under development’ versions so I do not know'', 45\% ``I am not familiar with ‘under development’ versions but feel possibly comfortable'', and 14\% ``I am familiar with ‘under development’ versions and feel possibly comfortable''. The 10\% ``other'' range from ``I'm willing'' to ``do not want viruses''.

\subsubsection{Phase~1 Summary and Key Findings}

Following recruitment and baseline survey, we conducted phase~1 of the study to explore how different motivational strategies might support older adults in becoming more physically active. This phase used familiar methods, such as SMS text messages, paper logs, and mailed materials. While this paper focuses on the app-based phase~2, we briefly summarize key aspects of phase~1 that shaped phase 2 and the development of \toolName{}.

During phase~1, 97 participants were randomly split into four intervention groups and a control group, and over 8 weeks received basic fitness-related instructions via bulk SMS. Non-control groups received additional motivational techniques (e.g., peer-support via our Facebook group). Participants logged their activities via paper forms or an online questionnaire and were encouraged to use Fitbit, Google Fit, or Pacer~\cite{pacerapp}.
 
A post-intervention survey (62/97 responses) and 54 semi-structured interviews (30--60 min) revealed positive outcomes. For participants with both (pre- \& post-intervention) survey results, average body mass index~(BMI) dropped from 28.1 to 27.6 in intervention groups (n=42), compared to a smaller reduction (29.3 to 29.1) in the control group (n=18). Among overweight participants (BMI 25+) this difference was more pronounced: 30.3 to 29.5 in intervention groups (n=30) versus 33.6 to 33.4 in the control (n=11).

Participants appreciated the structured exercise plan. One stated:

\begin{quote}
``This is a well-designed program. The exercises are easy to fit into my daily life, and I enjoyed the balancing exercises.''    
\end{quote}

Others reported physical and mental benefits, e.g.:

\begin{quote}
``I have experienced less pain from my Rheumatoid arthritis. Movement really helps.''
\end{quote}

On the flip side, some participants requested more challenging exercises, especially for hip and wrist mobility, and suggested gradually increasing difficulty. Participants also stated the following challenges and suggestions.
\begin{itemize}
    \item \textbf{Exercise tutorials that are easy to watch:} Participants suggested to add (and make it easy to watch) videos that demonstrate the exercises they were expected to perform. For example, one participant from the video-motivated group did not watch the videos as he feared that following a hyperlink from a SMS text message would take him off track from exercising.
    \item \textbf{Repetition flexibility:} Many participants considered fixed repetitions as too rigid.
    \item \textbf{Automatic logging:} A few participants found manual filling and mailing paper forms inconvenient, especially with deadlines. 
\end{itemize}

\begin{insight}
    While the phase~1 interventions worked as expected, participants voiced several opportunities for improvement that could be satisfied with a senior-specific fitness app.
\end{insight}

\subsection{Phase 2 Intervention With \toolName{} Phone App}

We invited all phase~1 participants to continue with phase~2. Of the 97 participants from phase~1, 25 chose to continue in phase~2. While not all participants gave specific reasons for opting out, we were able to confirm several common themes based on responses received via follow-up emails, text replies, and verbal comments during phone call check-ins. Several participants cited scheduling conflicts (vacations, weather, etc.), while a few mentioned illnesses, such as post-COVID fatigue or recent surgeries.

The six-month gap between the two phases and limited researcher communication during that time also contributed to disengagement. This suggests that future multi-phase interventions may benefit from running phases concurrently or in close succession to preserve momentum. Moreover, while we maintained contact through periodic updates, a more continuous communication strategy may help older adults feel more connected to the program and its goals. There were a few participants who had initial difficulty in downloading and understanding the app, and thus did not want to continue further.
These 25 participants (18 were female and 7 male, ranging in age from 65 to 85) formed the test group for phase~2.

\subsubsection{\toolName{} App Development}

Phase~1 feedback indicated that while bulk SMS and traditional motivational strategies were effective, participants desired additional features that go beyond traditional paper/SMS-based approaches. Smartphone apps are a flexible platform for delivering a variety of interventions, including
the ones we explored in phase 1. We thus built the \toolName{} app that implements both the traditional phase~1
motivational techniques (bulk text messages, peer support, etc.) and the improvement suggestions provided by
the participants (i.e., convenient video tutorials, automatic exercise tracking, etc.). 

\paragraph{\textbf{Design Considerations for Older Adults}}

We designed the \toolName{} app to accommodate the physical and cognitive changes associated with aging~\cite{liu2021mobile,kim2007contextual}. We thus aimed to (1)~simplify the interface to alleviate cognitive burdens and (2)~enlarge interactive controls to facilitate easier interaction~\cite{gomez2023design}. To simplify navigation, \toolName{} has a home screen that links to all main features. Each screen has a back button as a fallback mechanism~\cite{de2014design}. To reduce accidental touches, the app maintains generous spacing between interactive elements, following best practices that account for motor limitations often present in older users~\cite{de2014design,gomez2023design}.

To enhance usability, the app uses basic, distinct colors from a limited palette to prevent visual overload~\cite{liu2021mobile}. We increased the button and text sizes to make them usable even on the smallest devices~\cite{harte2017human,morey2019mobile} (and we tested the app on a wide range of screen resolutions). Specifically, any critical text is at least 36 point font size, and secondary text is at least 26 point. For colors we adhere to WCAG guidelines for high contrast ratios (contrast ratio of at least 4.5:1 for smaller text and at least 3:1 for larger text~\cite{w3c}. We prioritize text labels over icons. Any icon is concrete and semantically close to easily understandable symbols~\cite{footulcer,leung2011age,dosso2021older}.

To simplify installation and use, the user never has to enter a username, email address, or password. Instead, we use a simplified one-time login process~\cite{harte2017human} where, after installation the user only enters their first and last names and a contact phone number. The app uses this identifier to store workout data to \toolName{}'s cloud back-end without requiring further authentication.

From the phase~1 survey, we found that participants were split between Android and iOS device users. To ensure broad accessibility and reduce platform-specific development overhead, we implemented \toolName{} using Unity~\cite{unity}, a cross-platform mobile development framework. This allowed us to maintain a consistent user interface and feature set across both operating systems, while still accommodating device-specific capabilities such as GPS access and camera-based pose detection.

\paragraph{\textbf{Initial Deployment \& Key Features}}

At the start of phase~2, we released the \toolName{} app with a core set of features designed to support at-home physical activity for older adults. The app builds directly from the routines developed and the feedback received from phase~1. 

\toolName{} includes two primary modules for activity tracking: indoor exercises and outdoor walking. Each module is designed to support autonomous, self-motivated use. Unlike phase~1, participants were not given any laid out physical activity plan, instead, they were encouraged to engage with the app at their own pace and incorporate it into their existing routines.

The walking module uses the phone’s built-in GPS sensor to log walk duration and distance. The walk screen also shows brief motivational messages at five-minute intervals (e.g.: ``You can do it!''), and participants can start and stop walk tracking via on screen buttons. The first time a user initiates walk tracking, the app requests permission for location tracking. During a walking session, the app can run in the background and continue tracking.

The indoor exercise module includes 16~exercises, categorized into strength, cardio, mobility, and balance. We built the phase~2 exercise set based on the structured routines developed and distributed during the phase~1 intervention, which were informed by established public health guidelines~\cite{cdc, nih_nia}. These exercises were selected in collaboration with faculty and researcher from the Department of Kinesiology to reflect safe, evidence-based exercises appropriate for older adults.

To reduce the burden of manual logging, and address the need for clearer demonstrations, we integrated a real-time pose detection system using MediaPipe’s BlazePose library~\cite{bazarevsky2020blazepose}. This choice enabled hands-free exercise tracking, automated repetition counting, and visual feedback during exercise, all within a smartphone interface. This required the use of the phone's front-facing camera for exercise tracking.

BlazePose detects 33 key body landmarks with near real-time performance (30+ frames per second). Studies have shown that these marker-less body-tracking offer comparable accuracy to more expensive marker-based system~\cite{Wade22Applications, Latreche23Reliability}. We followed MediaPipe’s Android guidelines, adapting them for Unity for cross-platform deployment.

\begin{figure}
    \centering
    \begin{minipage}[b]{\linewidth}
        \centering
        \includegraphics[width=\linewidth]{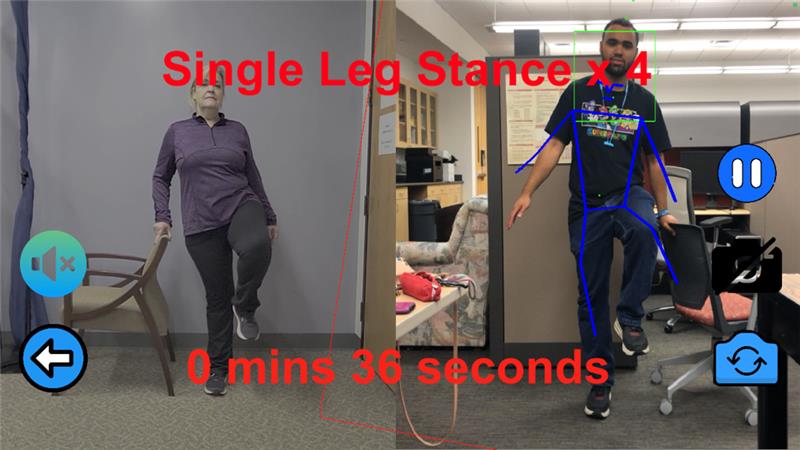}
    \end{minipage}
    \hfill
    \begin{minipage}[b]{\linewidth}
        \centering
        \includegraphics[width=\linewidth]{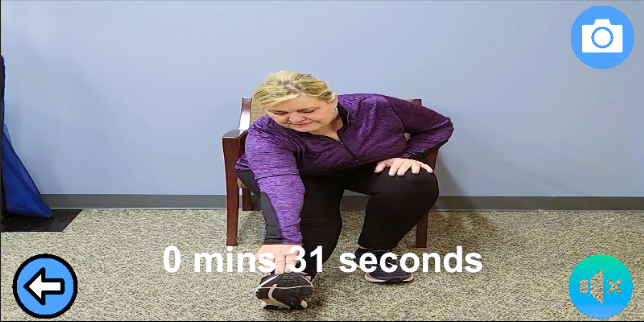}
    \end{minipage}
    \caption{Single leg stance detection via phone camera (top) and seated hamstring stretch without detection (bottom).}
    \label{fig:exercise}
\end{figure}

\begin{figure}[h!t]
    \centering
    \begin{minipage}[b]{0.3\linewidth}
        \centering
        \includegraphics[width=\linewidth]{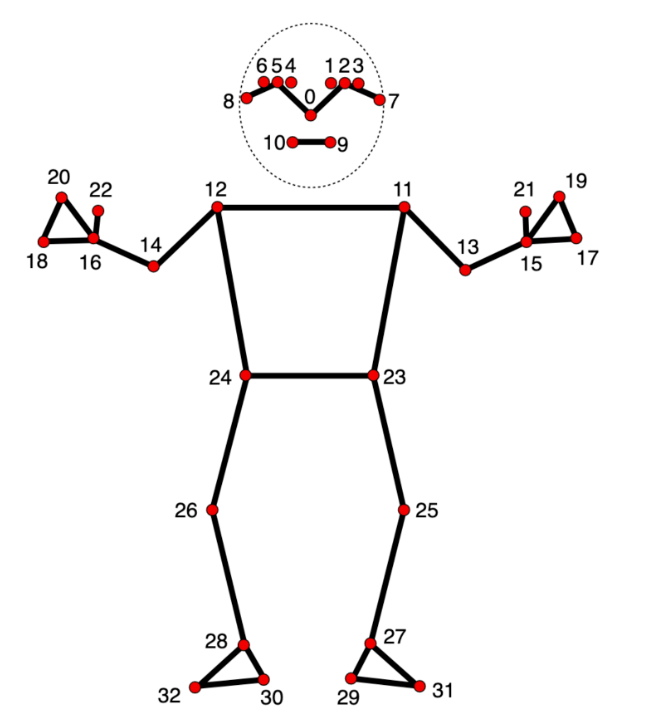}
    \end{minipage}
    \hfill
    \begin{minipage}[b]{0.3\linewidth}
        \centering
        \includegraphics[width=\linewidth]{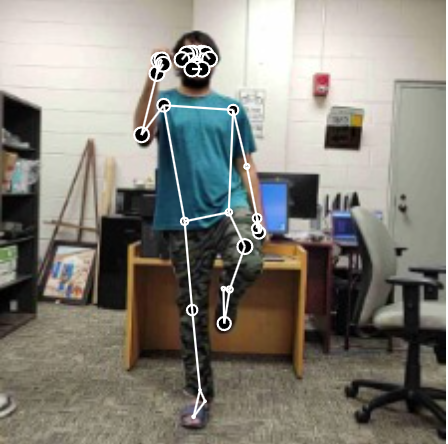}
    \end{minipage}
    \hfill
    \begin{minipage}[b]{0.3\linewidth}
        \centering
        \includegraphics[width=\linewidth]{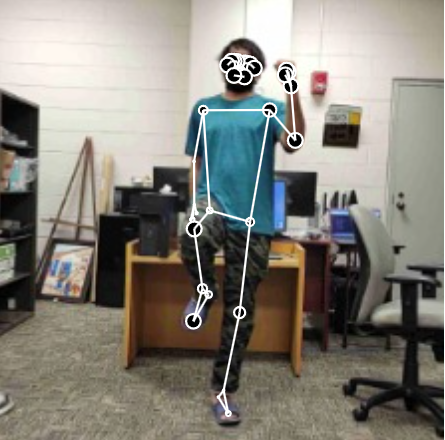}
    \end{minipage}
    \caption{BlazePose keypoints~\cite{bazarevsky2020blazepose} (left) and example body points detected by Blaze-pose during the team's training phase for the ``Single Leg Stance'' exercise.}
    \label{fig:blaze_pose}
\end{figure}

For exercise classification, we adapted MLKit's~\cite{mlkit} k-NN (k-nearest neighbors) code to match BlazePose-identified landmarks with pose samples from our training dataset. We collected 120 images per exercise in various lab conditions to account for diverse camera angles, environments, and body types. Figure~\ref{fig:blaze_pose} illustrates the detected body points. We encode each pose as a vector of normalized distances between body landmarks (as input for the k-NN algorithm). Although formal accuracy evaluation was outside this paper’s scope, pose classification was generally stable under varied conditions and was refined based on iterative testing.

Each exercise included two types of video demonstrations. First, during an active tracking session, the app visually guides users by displaying a short looping video demonstrated on the left, while the participant's pose-detection camera-stream appears on the right. The screen also displays the exercise name, duration, and repetitions the app detected (Figure~\ref{fig:exercise}, top). Second, the app offered a library of longer-form video tutorials that participants could browse at any time. These instructional videos provided verbal guidance, movement modifications, and safety tips. All videos were recorded by members of the Department of Kinesiology using a volunteer in her late fifties as the demonstrator. They were stored locally within the app to ensure offline accessibility.
 
While the app's main features are tracking walks and exercises, these are supported by several other features:

\begin{itemize}
    \item \textbf{Push notifications} sent every two to three days with motivational content (e.g., wellness reminders, links to external resources such as healthy recipe videos, stretching tips, etc.). This interval was chosen to balance encouragement with notification fatigue, aligning with persuasive design recommendations for older adults without becoming disruptive.
    \item A \textbf{written guidance section} to provide context about different activity types, intensity levels, and weekly recommendations, mirroring public health guidance for older adults.
    \item Built-in \textbf{peer support} linked directly to the Facebook group established during phase~1 (if the Facebook app is not installed, this opens the group in the default web browser). Rather than developing a new social platform, we chose Facebook because it remains the most widely used social network among older adults (\cite{socialmedia_pew}, see also Table~\ref{tab:digital_engagement_summary}). 
    \item An \textbf{in-app feedback form} to allow participants to report issues or suggestions directly from within the app.
\end{itemize}

\paragraph{\textbf{Mid-Study Adjustments}}

We adopted an iterative deployment model during the six-month Phase 2 study, releasing updates to the app approximately every two months. These updates responded directly to participant feedback and addressed emerging barriers to engagement. 

The first version of the app included all core features discussed above. All indoor exercises required the use of the camera for pose-based tracking, and activity history was not yet visible to participants. Several participants expressed difficulty positioning the phone to capture their movements or discomfort with camera-based tracking in general. In response, we added a non-camera mode for all exercises (Figure~\ref{fig:exercise}, bottom). This mode allows users to follow along with video instructions and self-pace through activities without activating pose detection. This update was released during the second iteration, with the goal of improving accessibility for those with privacy concerns or limited physical flexibility. 

Participants also reported uncertainty about whether their exercise efforts were being recorded and expressed the desire to access their activity report. Until this point, data was collected but not displayed in the UI. To improve transparency and reinforce engagement, we introduced a simple activity log in our third iteration that displayed completed exercises and walks, including date, duration, and repetition counts. 

\subsubsection{Local-first: Offline Use \& Pushing Data to Cloud Later}
Participants can exercise with the app independent of internet connection. Initially, the app saves data locally on the user's devices, including duration, date, and geographic coordinates for walks, and for indoor exercises, the exercise names, durations, and repetition counts. All other features are also available offline, including reviewing exercise logs and submitting feedback. When connected to the internet, the app securely uploads the newly stored data to our AWS DynamoDB database~\cite{aws_dynamodb}, ensuring no information is lost during network interruptions.

\subsubsection{User Training \& App Distribution}

To test \toolName{} in a real-world setting, we intentionally provided no in-person training sessions. Instead, participants received a manual containing step-by-step installation instructions and annotated screenshots. Technical support was available via phone and email. For Android distribution, the app was provided via a direct download link hosted on Google Drive, bypassing Google’s beta testing restrictions that require a Google account. For iOS users, the app was distributed through Apple TestFlight, allowing participants to install updates without requiring App Store approvals.

\section{Results: Phase~2}

We report findings from the six-month app deployment study (phase~2) involving ~25 older adults. We collected three types of data during this intervention: (1)~walk and exercise activity logs through the \toolName{} app, stored in the cloud database (2)~an exit survey in which participants rated their satisfaction with various app features on a scale from 1 (``strongly disagree'') to 5 (``strongly agree''). (3)~open-ended phone interviews, where participants shared their thoughts on their experience with the app. 
Each interviews lasted between 20 and 60 minutes, and were conducted over the phone. We transcribed and subsequently conducted a thematic analysis on the interview data, following a multi-stage coding process~\cite{vaismoradi2013content}. Two researchers independently performed initial open coding on the transcripts to identify recurring themes and patterns. The first author then reviewed the full set of codes, refined theme boundaries, and finalized a set of overarching themes through iterative synthesis.

\subsection{Participant Characteristics and BMI Change}

Based on our pre-study baselines survey, phase~1 participants have similar average BMI (28.9, n=97) as those continuing with phase~2 (28.7, n=25). Also similar are the shares of high-BMI (30+) participants, with 35\% (34/97) vs. 36\% (9/25), and living situation (Table~\ref{tab:combined_participant_data}). 

But phase~2 participants skew toward more tech-savvy (Table~\ref{tab:digital_engagement_summary}), formally educated, and physically active (Table~\ref{tab:combined_participant_data}). For example, 40\% of phase~2 participants hold advanced degrees, compared to 29\% in phase~1. Similarly, 54\% report walking at least three times per week  (vs. 29\% in phase~1), and 54\% engage in regular stretching (vs. 34\%). This difference indicates a self-selection of more educated, more physically active, and maybe more interested in working with university student researchers over long periods.

\begin{table}[h!t]
\centering
\caption{Participants' highest academic degree (top), living situation (middle), and pre-study activity levels (bottom) in phase~1 (n=97) and phase~2 (n=25). 
}
\begin{tabular}{lrr}
\toprule
(\%) & P1 & P2 \\ 
\midrule
No high school diploma & 4 & 0 \\
HS graduate or equivalent & 19 & 8 \\
2-year college degree & 12 & 12 \\
4-year college degree & 36 & 40 \\
Advanced degree(s) & 29 & 40 \\
\midrule
With spouse/partner & 40 & 40 \\
.. with children & 6 & 4 \\
.. with parents/close relatives & 3 & 4 \\
Living alone & 51 & 52 \\
\midrule
Walking \(\geq\) 3/week & 29  & 54  \\
Stretching \(\geq\) 2/week & 34  & 54  \\
Light strength training \(\geq\) 1/week & 25 & 33 \\
Moderate/heavy strength training \(\geq\) 1/week & 12 & 21 \\
Cycling \(\geq\) 1/week & 15 & 25 \\
\bottomrule
\end{tabular}
\label{tab:combined_participant_data}
\end{table}

As phase~2 lacked a control group (due to low recruitment), it is thus hard to directly compare results with phase~1 or draw definite quantitative conclusions from the  phase~2 exit survey. Broadly speaking, the average BMI of the 16 phase~2 participants (with reported BMI before and after phase~2) reduced modestly from 27.6 to 27.5. For the overweight subset (BMI 25+, n=11) the average BMI reduced from 30.0 to 29.7. Finally, for the obese subset (BMI 30+, n=5) the average BMI reduced from 33.9 to 33.2.

While these changes are encouraging, they are modest and exploratory. Due to the small sample size, absence of a control group, and possible self-selection bias, we do not draw statistical conclusions. However, the data suggest that participants are generally motivated and may benefit modestly from app-based support for physical activity.

\subsection{App Use and Survey Results}

\begin{figure}
  \centering
  \includegraphics[width=.8\linewidth]{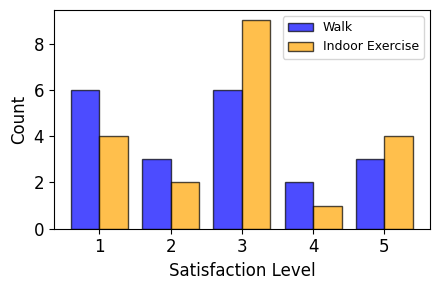}
  \caption{Participant satisfaction with \toolName{}'s tracking.}
  \label{fig:walk_exercise_satisfaction}
\end{figure}

App usage logs reveal diverse patterns of engagement with the \toolName{} app over the six-month intervention. Table~\ref{tab:participants_phase_2} summarizes the participant’s app usage, including indoor exercises and GPS-tracked walks. Half of the participants used the indoor exercise feature on fewer than seven days, while Sophia\footnote{All participant names are anonymized.}, the most engaged user, performed over 1k sets of exercises on 141 days. The most commonly performed exercise across all participants was shoulder-touch. The exit survey (Figure~\ref{fig:walk_exercise_satisfaction}) reflects mixed satisfaction with the indoor exercise tracking feature, with an average rating of 3.0 (SD = 1.4). 

\begin{table*}[htbp]
\centering
\caption{Participants (pseudonyms) with their self-reported age (as of June 2022), highest completed degree, and gender; 
S/I~=~participated in exit survey / interview; 
days with at least one app-tracked walk, 
months from first to last day of tracked walks,
total miles and daily average on tracked walks; 
days with $\ge 1$ indoor session, 
months from first to last day of indoor sessions, 
most-commonly used indoor exercise session type (with ties) and
their number, and 
sessions across all indoor exercise types.
HS~=~high school, incl. GED or equivalent;
Asc~=~associate degree (``e.g., 2-year college degree'');
4yr~=~4yr college;
Adv~=~advanced, incl. graduate and professional degrees;
Cs~=~chair sit to stand;
Mp~=~march in place;
Hs~=~Seated hamstring stretch;
Ls~=~single leg stance;
Sm~=~seated march;
St~=~shoulder touch.
}
\begin{tabular}{lrlcl|rrrr|rrrrr}
\toprule
 &  Age &  D & G & SI & \multicolumn{1}{c}{Walk} & \multicolumn{1}{c}{Rng} & \multicolumn{1}{c}{Total}  & \multicolumn{1}{c}{Avg} & \multicolumn{1}{|c}{Use}  &  \multicolumn{1}{c}{Rng}   & & & \\
& & & & 
  & \multicolumn{1}{c}{days} & \multicolumn{1}{c}{[mo]} & \multicolumn{1}{c}{[mi]}  & \multicolumn{1}{c}{[mi]} & \multicolumn{1}{|c}{days} & \multicolumn{1}{c}{[mo]} & Pop & Cnt & \multicolumn{1}{c}{All}\\
\midrule
Mary & 65 &4yr & f &S & 0 &&&& 3 & 2 & Ls & 4   &6\\
Thomas & 65 & Adv  &   m& SI & 4 & 0 & 0.1 & 0.0 &0\\
Diana & 66 &Adv  & f& SI &0 &&& &0\\
Samuel & 67 &4yr  &   m & I & 0 &&& & 13 & 7 & St &14   &32\\
Diane & 68 &Adv  & f &  \textendash   & 2 & 1 & 0.0 & 0.0 &  10 &   6 & Sm &   16  &  28 \\
Emily & 68 &4yr  & f& SI  & 19 & 6 & 19.2 & 1.1 &  16 &   9 & St &   42&  131\\
Evelyn & 68 & Asc  & f &S  &0 &&& &0\\     
Richard & 68 & Adv & m &\textendash & 0 &&& & 3 &   1 & St &       6   &18\\
Eleanor & 69 &Adv  & f& SI  & 22 & 1 & 19.0 & 0.9 &   1 &   0 & Ls & 2&2\\   
Georgia & 69 & Asc & f &S  & 3 & 0 & 2.7 & 0.9 &   2 &   0 & Hs   &   4   &    12\\
George & 70 &4yr  &   m &  \textendash  & 11 & 1 & 6.5 & 0.6 &0\\
Rebecca & 70 &Adv & f& SI & 17 & 5 & 19.1 & 1.1 &  12 &   6 & St & 22&   84\\
Marilyn & 71 &4yr & f& SI  & 116 & 5 &	61.4 &	0.5 &  18 &  10 & Cs,Mp & 22 & 71\\
Sophia & 71 &4yr  & f &S  & 42 & 5 & 21.7 & 0.5 & 141 &   6 & Ls &   292 &   1,190\\    
Peter & 72 &Adv &   m &S  & 0 &&& &   1 &   0      &       Cs &2 & 2\\
Anna & 74 &Adv  & f& SI &0 &&& &0\\     
Cora & 74 &Adv  & f &S  &0 &&& &0\\
Emma & 74 &HS & f& SI  & 64 & 7 & 41.7 & 0.7 &  61 &   9 & St &   120 &   463\\
Nina & 74 &4yr  & f& SI  &0 &&& &   3 &   4 & St &   10   &24\\
Scott & 74 &4yr  &   m & I & 89 & 6 & 69.5 & 2.1 &0\\
James & 75 &4yr  &   m &S & 16 & 2 & 2.7 & 0.2 &  19 &   3 & Cs &17     & 40\\    
Julie & 80 &Adv  & f& SI  & 94 & 10 & 7.9 & 0.1 &   5 &   2 & St &   12  & 30\\
Joan & 81 &HS & f &S &0 &&& &   3 &   1 & St &   12  &   24\\
Clara & 83 &4yr & f& SI  & 16 & 3 & 2.0 & 0.1 &  24 &   4 & Ls & 60 & 148\\
Rachel & 85 & Asc & f& SI & 41 & 3 & 36.8 & 0.9 &   6 &   8 & Cs,Ls & 6 & 24\\    
\bottomrule
\end{tabular}
\label{tab:participants_phase_2}
\end{table*}

The GPS walk data similarly shows a wide range of usage patterns. Five participants logged only short walks, averaging 0.2~miles or less per walk, over spans ranging from 2 to 94 days. Nine participants logged moderate walk distances, averaging 0.5–1.1~miles per walk over 3 to 116 days. One outlier (Scott) recorded an average of 2.1~miles per walk. Some short-walk participants, like Clara and Julie, used the app on indoor exercise equipment, yielding noisy ``short-walk'' GPS readings. Specifically, Clara used her treadmill, and Julie (due to a medical condition) used an elliptical machine. According to the exit survey, both satisfaction and likelihood of recommending the app for walk tracking averaged 2.7 (SD 1.5) on a 5-point Likert scale (Figure~\ref{fig:walk_exercise_satisfaction}).

Additional survey responses highlighted varying levels of satisfaction with other app features. Relatively highly rated, participants found it easy to interact with the Facebook group (avg. 3.8, SD 1.5). Video tutorials were mostly considered helpful (avg. 3.6, SD 1.2), followed by activity guidelines and notifications (both avg. 3.5, SD 1.2 and 1.3) and app navigation (avg. 3.4, SD 1.3). In-app instructions received the lowest rating (avg. 2.8, SD 1.4), indicating challenges with understanding how to use the app effectively.

In summary, some participants used the app regularly and found certain features helpful, while others faced usability barriers that limited their engagement.

\subsection{Comparing With Our Paper-Based Intervention}

Participants expressed a range of preferences when comparing the \toolName{} app to the paper and SMS-based intervention used in phase~1. On the pro-app side, participants appreciated the \textbf{convenience of the app for tracking progress} without needing traditional mail-based methods: 

\begin{quote}
``I liked the paper too. But I knew that that wasn't cost-effective even for you guys, you know, the mail that up every time. Or as you know, the app could just track me.'' -- Emma
\end{quote}

Participants found the app comprehensive, covering various aspects of physical exercises and providing valuable information: 

\begin{quote}
``The exercises cover all the bases. The information is spot on. I find it geared to people who don't go to the gym or use equipment, which is fine.'' -- Samuel
\end{quote}

Samuel also noted his appreciation for the social aspect of working out in gym settings: 

\begin{quote}
``I like working out in gyms and classes for a bit of social activity. I workout better around others. But not everyone can get out and about. Well rounded program. Movement is critical as we age.'' -- Samuel
\end{quote}

On the flip side, Clara preferred the structure of phase~1, deeming it better suited to her needs. She (and several others) found the \textbf{app challenging to navigate}. Marilyn expressed a preference for filling out paperwork. Rebecca echoed this sentiment, favoring the pencil-and-paper method because it held her accountable, stating: 

\begin{quote}
``I like the pencil and paper, at least you were being accountable. You know, I know, I knew I had to count everything...And you had to send it in, I think, every week. So you were really focused on doing it because you knew you had to send it in.'' -- Rebecca
\end{quote}

Rachel’s reflections further highlight this tension between app benefits and the appeal of simpler, more structured methods. She appreciated how app notifications prompted her to reflect: 

\begin{quote}
``I think every time that it reminded me of something, there would probably be some minimal benefit there you know, if you put a comment out about what do you do to de-stress and I had to think about that a little bit and realize what was the main thing ... the main thing I do is go for walks in nature.'' -- Rachel
\end{quote}

Yet, she also voiced frustration with the app's usability and expressed a preference for a simpler setup like phase~1, which she recalled as being more straightforward, which provided her with the sense of making progress. 

\begin{insight}
Participants appreciated the app’s convenience and breadth of features, but many also missed the structure and accountability of the paper-based phase~1 approach. 
\end{insight}

\subsection{Balancing Activity Tracking Automation and Control}

Participants' experiences with tracking walks and indoor exercises revealed diverse preferences about how their physical activity should be tracked, ranging from fully automated methods to manual input.

Some participants found the requirement to manually start and stop walk tracking intuitive. Emma, who logged 64 walk days over seven months with an average of 1.1 miles per day, compared it to using a stopwatch. She praised the simplicity of the feature: 

\begin{quote}
``Once I clicked on it, it was easy. It was like putting a stopwatch on and once you completed the walk, and then you stop stopwatch.'' -- Emma   
\end{quote}

Similarly, Julie, who logged 94 days over 10 months, found the feature helpful: 

\begin{quote}
``the app just kind of lets me know I’ve reached my 30 to 35 minutes. So that part I liked, ... I could probably do that on the Fitbit. But I don’t have one.'' -- Julie
\end{quote}

Marilyn, who logged 116 walk days over five months, told us: 

\begin{quote}
``I used it for when I was walking...every time I made it automatic deal. Go and start the walk.'' -- Marilyn
\end{quote}

However, others found manual tracking inconvenient. Rachel, who logged 41~walk days over three months (avg. 1.5~mile/day), shared: 

\begin{quote}
``I didn’t find that the walking aspect on your app worked as well for me and because partly, I couldn’t remember to tell it when I was finished. If I started fine. And then I would remember hours later that Oh, I didn’t tell it.'' -- Rachel
\end{quote}

This oversight led to inaccurate distance tracking. To avoid large distance over-measuring (e.g., including a post-walk car ride), we manually checked all walk entries and only kept walks with sub 4~MPH average speed.

Scott, who logged 89 walk days over six months (avg. 3.3~mile/day, which reflects his goal of covering 10k steps on 3+ days per week), said: 

\begin{quote}
``The thing I really like about the Garmin watch is that I don't have to remember to turn it off and turn it on in my exercise program or whatever I'm doing. It automatically tracks. Okay, I know you're, you're probably trying to develop your own app for your own purposes. And I get that I understand that for me, if it were integrated with something that I use already, I wouldn't have to worry.'' -- Scott
\end{quote} 

Although he understood the research purpose of the \toolName{} app, he preferred using his Garmin watch for automated tracking. Similarly, Anna noted: 

\begin{quote}
``I like something that tracks things automatically, like for example, you know, best I can do that with my phone or my Apple watch, or you know, those kinds of things. It's kind of a pain to have to go to the app and say, Okay, I'm gonna, I'm gonna walk down, but I have to remember to stop. I kinda like to be automated.'' -- Anna
\end{quote}

Emily echoed these concerns: 

\begin{quote}
``If I walk or if I do exercise, Google Fit automatically records that you're moving. But your app, I'm sorry, you have to open it up. And so I forget to open it up when I'm exercising.'' -- Emily
\end{quote}

While several participants expressed a preference for automatic tracking via devices like Garmin or Apple Watch, we intentionally designed the \toolName{} app to function independently of external wearables to ensure accessibility, reduce cost barriers, and maintain simplicity. 

\begin{insight}
Some participants prefer manually starting and stopping activity tracking, while others prefer more automation. Some participants would like to incorporate their existing fitness-related hardware.
\end{insight}

\subsection{Clear and Motivating Feedback}

Participants had mixed experiences with the app’s feedback mechanisms. Several valued the built-in exercise demonstrations and notifications, which helped them maintain proper form and stay active. Emma, who used the app for nine months, noted: 

\begin{quote}
``It reminded me to keep doing the exercise and the workouts, the stretching, and stuff like that. That’s what I like the best.'' -- Emma
\end{quote} 

Similarly, Emily emphasized the benefit of clear demonstrations: 

\begin{quote}
``I would say the benefit would be in having the demo of how to do the particular exercise.'' -- Emily
\end{quote}

Other participants echoed that the indoor exercise features supported flexibility and encouraged activity, especially when outdoor walking was not an option.

At the same time, a few participants highlighted the lack of clear feedback on their exercise performance as a major barrier to engagement. For example, Clara used the app for four months but had a relatively low engagement with the exercise feature. She felt unsure about whether she did it correctly: 

\begin{quote}
``I could use the camera. But I evidently didn't realize that I was doing something wrong.'' -- Clara
\end{quote}

Similarly, Rebecca described confusion over the exercise tracking process: 

\begin{quote}
``I did go into the app, and I used it...[Were] they then recorded, or were they supposed to be able to give me feedback or anything? I never knew that. I never got any feedback...I tried the exercises...I didn't get any summary of what was being recorded or anything.'' -- Rebecca
\end{quote}

Initially, for indoor exercises, the app offered only real-time feedback through visual overlays, i.e., lines joining body keypoints, along with a repetition counter and duration display. These visual cues were intended to help users understand whether they were performing the movements correctly, as the repetition count would only increase upon proper execution. However, this feedback was limited to the active session. Once the session ended, there was no summary or confirmation screen, leaving users uncertain about whether their activity had been recorded. This absence of post-session feedback caused confusion and, for some, reduced motivation to continue.

Later in the intervention, we introduced an \textbf{\textit{Activity Records}} feature, allowing users to view their exercise history. Since this update came mid-study, some participants may have already lost interest or motivation due to earlier confusion or unmet expectations. Thomas, who had minimal engagement, reflected this frustration: 

\begin{quote}
``At first, it didn’t seem to keep the results.'' -- Thomas
\end{quote}

Technical limitations added further confusion. Because GPS-based walking logs did not capture treadmill or elliptical workouts accurately, some users saw near-zero distance readings. Clara, for instance, experienced this while using her treadmill and remarked: 

\begin{quote}
``I couldn't tell if it was working or not... it seems like if it was working, it was only collecting data and not giving me any information.'' -- Clara
\end{quote}

\begin{insight}
Participants like both immediate and post-session feedback, reassuring them that their activity was recognized and recorded, as well as tools to reflect on their progress over time. When such feedback was absent or delayed, trust in the system weakened, and motivation declined.
\end{insight}

\subsection{Accessible and Embedded Social Support}

Social and peer support can significantly enhance engagement with digital health tools among older adults, offering emotional encouragement and accountability. To facilitate this, the \toolName{} app reused the phase~1 Facebook group, where researchers actively posted weekly motivational prompts and questions designed to encourage participant engagement and community-building. Some participants used the platform to share their experiences. Emily posted photos and videos of her walks to document her \toolName{} app use. Others shared updates on their health, such as post-surgery recovery or experiences related to COVID-19, which helped participants relate to one another and build a supportive community. But the overall engagement remained limited.

Diana found the group motivating, saying: 

\begin{quote}
``the Facebook group. I had a good experience with that. I did think it was kind of an incentive as well, to use the app and to you know, do the exercise.'' -- Diana
\end{quote}

She also noted that participation was sparse: 

\begin{quote}
``My only disappointment with the Facebook group was that I don't know how many participants you had overall, but not many of them were in the Facebook group or were active. That was the only problem that I had with that. I think that it would have been it was nice, because you all were with your group was was interacting and posting stuff pretty regularly. That was an incentive to stay involved.'' -- Diana
\end{quote}

Many participants reported minimal or no engagement due to personal preferences or platform barriers. Samuel mentioned: 

\begin{quote}
``I'm not on Facebook, which cuts that out for me. Not being on Facebook, I feel, isolated me.'' -- Samuel
\end{quote}

Thomas echoed this, suggesting: 

\begin{quote}
``I'm not saying you should restrict it to only people on Facebook. But if there would be a way to have a community outside of that...because I felt like I was interested, I tried it out. But I had no feedback at all, basically, except from the staff. So it's something to consider as you go to the next page or next time is, how do we get results and help the community see results from people who aren't on Facebook.'' -- Thomas
\end{quote}

These experiences highlight how reliance on an external platform can unintentionally exclude older adults and create accessibility barriers.

Even among users comfortable with Facebook, consistent reciprocal interaction was limited. Rachel, who actively attempted to stimulate interaction, vividly described these challenges: 

\begin{quote}
``And I tried to participate somewhat in the Facebook page...give you some feedback and also help generate some interest with other people. I sense that maybe a lot of people being old, don't really use Facebook that much or weren't as comfortable getting involved...noticed that that was pretty quiet. Even though you guys were trying, you put out the the teasers and the challenges and suggestions. And that was good. That was helpful.'' -- Rachel
\end{quote}

Other participants also acknowledged their minimal participation. Emma admitted: 

\begin{quote}
``I only comment like two or three times... So I didn't do a whole lot of Facebook...that's just not me.'' -- Emma
\end{quote}

Similarly, Julie remarked:

\begin{quote}
``I do comment on that periodically... We didn't really talk back and forth.'' -- Julie
\end{quote}

\begin{insight}
While the community support goal was appreciated, relying on an external platform created barriers, i.e., some participants were unfamiliar with or resistant to Facebook. Others found limited engagement from peers demotivating. 
\end{insight}

\subsection{Content Flexibility and Routine Alignment}

Participants emphasized that digital fitness tools must adapt to their personal routines and preferences, not the other way around. Many participants described a disconnect between their daily exercise routines and what the app expected them to do. 

Some participants noted that the app’s exercise offerings did not align with their regular workout routines. Thomas, for instance, shared: 

\begin{quote}
``I did try the app for a little bit at first. But it didn't correspond with what my main workout stuff was.'' -- Thomas
\end{quote}

Another participant, who worked out with a personal trainer, commented that the app was not designed for users at their activity level, and thus, didn't use the app. 

Several participants called for more variety in the app’s content. Rebecca suggested that repeating the same routines for extended periods reduced engagement: 

\begin{quote}
``change it up? Okay, yeah. If I was doing this thing, I would change it up every week, I wouldn't have been doing the same thing for months. I would put different exercises up there.'' -- Rebecca
\end{quote}

Rather than following a fixed program, many participants preferred adapting the app's features to fit their preferences. Emily shared how she incorporated her own variations: 

\begin{quote}
``I did variations ... like arm curls before getting out of bed, marching when I’m seated ... the ones you have in there, I think I did those kinds of things, too.'' -- Emily
\end{quote}

Even among those who were interested in structured activity, the app didn’t always meet their expectations for depth or challenge. Rachel explained that she struggled both with the physical setup and with the simplicity of the exercises:

\begin{quote}
``I have found it awkward to use. To try and prop the phone up to see my movements. Also the exercises I saw are too simple to hold my interest.'' -- Rachel
\end{quote}

Participants also pointed out that much of their physical activity took place outside the app, through online classes, gym visits, or independent walking routines, but the app offered no way to log or acknowledge these efforts. This lack of integration limited the app’s role as a trusted tool for tracking and reflection. One participant explained: 

\begin{quote}
``I did some online exercise classes with  Methodist Hospital and with SilverSneakers on Zoom...so you know, I'm staying active, I just have no way of recording it to the app. It might be nice that, you know, since we don't have the paper that we could go into seeing your Fit app and put in what we've done.''
\end{quote}

Some participants used the app only for specific exercises that aligned with their needs, and disregarded the rest. As Samuel put it: 

\begin{quote}
``I referred to it from time to time mainly about balance exercises.'' -- Samuel    
\end{quote}

\begin{insight}
Participants value flexibility over rigidity, often adapting exercises to their own routines, using the app selectively, or expressing frustration when the content did not align with their established habits.
\end{insight}

\subsection{Privacy Concerns and Confidence in the App's Data Collection Practices}

User privacy was a key consideration in \toolName{}'s development. Several participants raised early concerns about the app's access to the phone's camera for indoor exercise tracking. Some chose not to continue with the study due to these concerns. For example, Dorothy said: 

\begin{quote}
``I am not participating in phase 2. I don't like the idea of the app turning on my camera. The idea is too invasive.'' -- Dorothy
\end{quote}

Others echoed this discomfort. Georgia shared her mistrust of the camera feature and chose to only use the outdoor walking feature. Irene stated: 

\begin{quote}
``I am still considering whether to use the app. I don't know if I want to be tracked. I would be interested with no camera option.'' -- Irene
\end{quote}

Even participants who continued with the app expressed early hesitation. Samuel said: 

\begin{quote}
``I'm interested in staying with the program but was concerned with its access to my iPhone. If it's like a normal app, I need the link.'' -- Samuel
\end{quote}

After the researchers clarified what data the app would collect, Samuel chose to install it and used the exercise features. However, these reassurances were not always sufficient to eliminate concern.

\begin{table}[h!t]
\centering
\caption{Participant perceptions on data collected by the \toolName{} app: 
``strongly disagree'' (1) to ``strongly agree'' (5), 
n=20;
standard deviation (SD) 1.4 each.}
\begin{tabular}{lc}
\toprule
Question & avg \\
\midrule
Understand the types of data collected & 2.8 \\
Understand the amount of data collected & 2.4 \\
Comfortable with the collected data & 3.0 \\
Feel in control of the collected data & 3.0 \\
\bottomrule
\end{tabular}
\label{tab:data_privacy}
\end{table}

In response to the camera-related concerns, we added a non-camera option for indoor exercises and informed participants that the app performed all processing locally without storing or transmitting photos or videos. Despite these measures, participants reported only partial confidence in the app’s data practices. Survey responses (Table~\ref{tab:data_privacy}) showed low-to-moderate levels of understanding and comfort: on a five-point scale, participants rated their understanding of the types of data collected at 2.8 and the amount of data collected at 2.4. Comfort with data collection averaged 3.0, as did feelings of control over the data.

\begin{insight}
Participants are concerned about data privacy, especially the app's use of the in-phone camera for indoor exercise tracking.
\end{insight}

\section{Discussion}
Older adults make up a growing portion of the global population, and many are increasingly turning to digital tools to support their daily lives, including health and wellness~\cite{techuse}. While physical activity has traditionally been supported through structured, non-digital interventions such as paper logs or in-person programs, there is now a growing shift toward more flexible, technology-based approaches that can be used independently. But when older adults are offered both a structured, paper-based intervention and a self-directed smartphone app, which do they prefer? And what shapes their choices?

Findings from the \toolName{} deployment reveal both the promise and complexity of standalone fitness apps for older adults. While some participants engaged with the app and appreciated features like video guidance and motivational reminders, others struggled with usability challenges, mismatched routines, or discomfort with camera-based tracking. These mixed experiences highlight key design tensions between autonomy and structure, automation and control, privacy and feedback. Below, we discuss these themes in relation to prior work and consider implications for accessible fitness app design for older adults.

\subsection{Designing for Flexible Automation and Control}
One of the key design tensions we found was balancing manual control with automatic tracking. Some participants valued the app’s manual start–stop design, describing it as intuitive and comparable to using a stopwatch. They appreciated being able to decide when to begin and end an activity, which reinforced a sense of control. Others viewed manual tracking as a burden, noting they often forgot to stop recording and wished that the system would log walks automatically. For participants accustomed to devices like Garmin or Apple Watch, the lack of automation felt like a step backward in convenience. This challenge mirrors findings from prior works, while manual controls can support privacy and control, many older adults find them cumbersome and prefer more passive tracking, often via wearables~\cite{harrington2018designing}.

Prior studies suggest that most wearable activity-tracking systems for older adults mostly focus on passive monitoring of walking and falls, often overlooking broader exercise or other activity support~\cite{steinert2018activity,vargemidis2020wearable}. For \toolName{}, we deliberately avoided wearable integration, not just to lower costs and reduce setup burden, but also because we wanted to support camera-based exercise tracking. A goal of our app was enabling older adults to count exercise repetitions and receive feedback using only their smartphones. Introducing wearables would have required two separate tracking systems, potentially complicating the experience. While this design reduced barriers, it also meant that participants already comfortable with wearables missed the automation they expected.

\noindent\textbf{Recommendations:}  
\begin{itemize}
    \item Support hybrid approaches, i.e., offer a lightweight manual option for those who value control and simplicity, while also offering integration with popular wearables for those seeking automation. 
    \item Provide clear indicators of when tracking is active or paused to reinforce trust and avoid uncertainty.
\end{itemize}

\subsection{Designing for Adaptability and Personalization}

Our findings indicate that older adults want fitness apps that can flexibly integrate into their exercise routines; they are not seeking rigid exercise options. While \toolName{} provided both walking and exercise modules, many participants felt the routines were too uniform or mismatched with their daily lives. Several participants engaged selectively with the app, choosing only the features that aligned with their existing routines, while others disengaged entirely when the app failed to meet their expectations for challenge or variety. This echoes prior studies, showing that adaptation and customization are essential in promoting engagement among older adults~\cite{kappen2018gamification}. Customization in fitness apps provides a sense of autonomy to the users, which in turn strengthens adherence and enjoyment~\cite{bol2019customization}.

As a form of customization, \toolName{} included an Activity Guidelines feature offering information on warm-up and cool-down, exercise intensities, and weekly recommendations. This was designed to give participants the freedom to choose how to engage with the app, rather than imposing a single fixed program. However, these guidelines were static, remaining the same throughout the six-month deployment. While participants appreciated having the tools available, several suggested that more dynamic support, such as changing weekly routines, difficulty adjustments, or integration with non-app activities, would help with motivation. \toolName{}, being a lightweight, proof-of-concept prototype, provided only a limited set of exercises, which failed to meet participants' expectations of variety.

Features such as weekly content rotation, manual input for non-app activities, or adaptive suggestions based on user history could foster deeper engagement and a stronger sense of ownership.
Feedback suggested that this rigid, one-size-fits-all approach often felt mismatched with their lifestyles, especially for those with pre-established habits.
Older adults are not seeking rigid programs. They want tools that can adapt to their lives, support a range of activity levels, and acknowledge both structured and informal physical activity.

\noindent\textbf{Recommendations:}

\begin{itemize}
    \item Provide broader exercise selection, including different variations, difficulty levels, and equipment-specific adaptations to meet diverse needs.
    \item Introduce weekly variations, seasonal challenges, or adaptive suggestions based on prior use to prevent monotony.
    \item Introduce gradual increases in challenge or variety based on prior use, helping participants maintain engagement without overwhelming them.
    \item Enable manual logging of non-app activities (e.g., gym classes, gardening, swimming), so informal routines count too.
    \item Offer lightweight personalization (e.g., choosing preferred exercise categories) while keeping defaults simple for those who value consistency.

\end{itemize}

\subsection{Motivation through Guidance and Feedback}

Our findings highlight the importance of combining guidance and feedback to sustain motivation and trust in digital health tools for older adults, in line with previous studies~\cite{yfantidou202314}. Participants valued the app's instructional support in the form of video demonstrations and timely notifications. It helped them maintain proper form and reminded them to stay active, even when outdoor activity was not possible. 
At the same time, the gaps in the app's feedback design limited participants' sense of progress. Several users described uncertainty about whether their efforts were being recorded correctly, or whether they were performing movements as intended. While \toolName{} provided real-time cues like repetition counters or duration of sessions, it did not deliver post-session activity summaries until after deployment. Without these cumulative views, participants often felt their efforts went unnoticed, which weakened trust in the system and reduced motivation to continue.

These experiences align with broader evidence from self-tracking research. While real-time feedback is useful, it is rarely sufficient on its own. Cumulative feedback validates user effort and makes progress visible over time. Simplified and accessible feedback, combined with praise, rewards, reminders, and suggestions can build user confidence and encourage long-term adherence to physical activity among older adults~\cite{harrington2018designing,yfantidou202314}.

\noindent\textbf{Recommendations:}

\begin{itemize}
    \item Pair instructional guidance with post-activity acknowledgment to ensure users feel both supported in the moment and validated over time.
    \item Include end-of-session summaries and weekly progress reports to support reflection.
    \item Design feedback that is simple and interpretable, using clear visuals, icons, or single progress numbers to reduce cognitive load.
    \item Incorporate praise or lightweight rewards (e.g., congratulatory messages, icons) to reinforce positive behaviors.
    \item Provide timely notifications that act as gentle reminders without overwhelming users.
\end{itemize}

\subsection{Inclusive Peer Support}

Social support is a well-established determinant of health and well-being in later life. For older adults, strong social ties can improve physical health and overall quality of life~\cite{chen2014loneliness}. Social networking sites extend the potential for such support by offering accessible spaces for encouragement, accountability, and community building, even when in-person connections are limited~\cite{chew2023untangling}.

We used a dedicated Facebook group to deliver peer support, but our findings reveal both potential and limitations. While some participants described the Facebook group as motivating, citing researcher posts and occasional peer updates as encouragement, many others engaged little or not at all. Several participants were not Facebook users and therefore felt excluded from the community altogether. Even among participants who did use the group, participation was often minimal and one-sided.

These experiences point to a broader accessibility issue: when social support relies on an external platform, it risks excluding users who are unfamiliar with or resistant to that technology. Though Facebook use have shown to provide a sense of belonging among older adults, many remain hesitant due to concerns about privacy, lack of interest, and limited digital literacy. Most participants in our study mainly reported using Facebook only to view photos or posts from friends and family, without deeper interaction. This passive participation created negative experiences for those who wanted to connect more actively. This suggests that while social connection via social networking sites can foster perceived support, its effectiveness depends on familiarity, perceived usefulness, and active, reciprocal engagement. For older adults with varied digital literacy and interests, platform choice can unintentionally lead to exclusion rather than connection, reinforcing prior findings that technology decisions can significantly impact older adults' accessibility and engagement~\cite{xing2024designing}.

\noindent\textbf{Recommendations:}  
\begin{itemize}
    \item Embed social support directly within health apps, rather than relying on third-party platforms.
    \item Provide lightweight, low-barrier interaction options (e.g., in-app message board, photo-sharing space, or activity leaderboards) that don't require separate accounts.
    \item Support engagement with prompts or features that encourage reciprocal participation.
\end{itemize}

\subsection{Communicate Privacy Protections Clearly}

Even with careful efforts to minimize data collection and keep it transparent, such as not storing video, and offering a non-camera option, several participants expressed concerns or confusion about what the \toolName{} app recorded. Privacy discomfort was a central reason some chose not to participate in phase~2, particularly due to the use of the camera for indoor exercise tracking. Others who did use the app reported only moderate comfort and limited understanding of what was being collected.

This points to an important tension when designing for older adults. Technical privacy protections are often not enough if users don’t feel informed or in control. Berridge et al. found that older adults’ trust in elder care technologies increased when they felt in control, not just over what data was collected, but how that data was used and communicated~\cite{berridge2022control}. In their study, over 90\% of participants rated the ability to pause a system as very or extremely important. In \toolName{}, we attempted to support control through manual tracking (e.g., starting/stopping walks) and a non-camera interface, yet our findings suggest these measures were not always interpreted as privacy protections. 

Prior research shows that older adults often misinterpret or make assumptions about data flows, especially when interacting with sensor-based or Artificial Intelligence technologies~\cite{frik2019privacy}. Unclear data handling often contributes to limited engagement~\cite{quinn2019mobile,kaur2022usability}. Although we reassured participants that \toolName{} did not save any photos or videos, and that all processing occurred locally, these verbal assurances, or notes during onboarding, were insufficient to build lasting trust. 

\noindent\textbf{Recommendations:}
\begin{itemize}
    \item Provide persistent in-app reminders about what is and is not being collected, not just one-time explanations.
    \item Offer simple, accessible tools such as a FAQ or a privacy dashboard that summarizes what data is recorded and why.
    \item Give users visible, easy-to-use controls to pause or disable features, reinforcing their sense of autonomy and choice.
\end{itemize}

\subsection{Limitations}

Our study presents several limitations that should be considered when interpreting our findings. First, our participant recruitment was shaped by both self-selection and technology requirements. Our sample primarily included individuals who were already interested in physical activity and willing to try a new app, which may not reflect the broader diversity of older adults. Moreover, phase~2 participants were predominantly women, and the overall sample size was modest, factors that limit generalizability.

Second, technical constraints affected both recruitment and app use. Several older adults were unable to participate due to compatibility issues with Android version 6 or lower, underscoring how device limitations can create barriers to inclusion. Among those who used the app, features like indoor exercise tracking required careful phone placement, which some participants found frustrating. Occasional inaccuracies in repetition counting further impacted overall engagement and trust in the system. Since \toolName{} is a prototype, its limited exercise variety and detection precision may have further contributed to participant dropout or selective usage.

Third, a six-month gap between phase~1 and phase~2 disrupted study continuity and made it difficult to compare participant experiences across phases. Some participants became unavailable, while others may have experienced unrelated changes in motivation, routines, or health. Future studies should aim to evaluate interventions in parallel—ideally with a control group, to better isolate the effects of the app itself.

Lastly, we did not screen for existing use of fitness technologies. Some phase~2 participants were already relying on wearable devices or structured exercise programs, which reduced their need for the app and likely contributed to lower engagement. Future work should consider how digital tools complement or compete with older adults’ existing routines and technologies.

\section{Related Work}

Over time, digital interventions have become a promising alternative to traditional, paper-based, or in-person fitness programs for older adults. These technologies vary widely in their design and delivery, ranging from simple step counters to more complex systems using smartwatches, pose estimation models, or even augmented/virtual reality (AR/VR)~\cite{cadmus2015randomized,chen2021development,chen2024silvercycling}. 

Many digital interventions rely heavily on wearable devices, such as pedometers, fitness trackers, smartwatches, etc. that passively log steps, heart rate, or energy expenditure~\cite{cadmus2015randomized,li2024electronic}. These tools can provide a low-effort way to monitor physical activity, but they also introduce a number of barriers. For one, they often require setup steps that assume technical fluency, such as syncing devices via Bluetooth, configuring settings, frequent charging, etc.~\cite{moore2021older}. For older adults with limited experience in configuring mobile devices, these requirements can be intimidating or frustrating. Besides, a significant portion of wearable technologies focus on supervising older adults (e.g., fall detection, adherence monitoring) rather than supporting them in self-directed engagement with physical activity~\cite{vargemidis2020wearable}. Design limitations also contribute to limited adoption. Because older adults are rarely involved in the early stages of development, many systems overlook the diversity of real-world needs and often elicit negative emotional responses~\cite{vargemidis2021irrelevant}.

Even in more exploratory systems, like exergames, which combine physical activity with interactive gaming elements,  or AR/VR-based interventions, researchers frequently rely on external hardware, such as depth sensors, motion-capture setups, installed in particpants' homes or tested in lab-based simulations. For instance, Ogonowski et al. deployed an ICT-based system combining exergames with continuous monitoring, explicitly aimed at fall prevention in older adults’ homes~\cite{ogonowski2016ict}. Their six-month ``Living Lab" study demonstrated positive outcomes in self-perception and fall-awareness. Similarly, Uzor and Baillie evaluated a tailored exergame system called Recov-R for home-based fall risk reduction, conducting an 8-week randomized controlled field study with 38 participants~\cite{uzor2019recov}. Their findings indicated improved exercise adherence, and balance among older adults. However, both relied on specialized sensors and installation by trained personnel. While feasible in controlled deployments, these requirements pose clear barriers to scalable, unsupervised use in everyday settings. Similar trade-offs exist in immersive AR/VR interventions designed to promote physical activity~\cite{eisapour2020participatory,mostajeran2020augmented,blomqvist2021using}. While they have potential for targeted outcomes, these systems often require significant infrastructure and supervision, and are less practical for autonomous use by independently living older adults.

In parallel, mobile and tablet-based interventions have emerged as a more lightweight alternative, avoiding the setup barriers of wearable or immersive systems. Recent advances in pose estimation and movement tracking are being explored to deliver guided exercise routines using only a device’s built-in camera and on-device computation, without requiring external hardware. Using computer vision and machine learning, these systems analyze body movements and provide real-time feedback, supporting accuracy and reducing injury risks~\cite{o2013current,yao2023poserac}. For instance, Luo et al. has explored the complexities in fall detection and proposed a pervasive pose estimation scheme for fall detection~\cite{luo2022pervasive}. Chen and Or developed an ML-based exercise system for older adults with chronic knee pain, combining video demonstrations with real-time feedback and performance scoring~\cite{chen2021development,chen2023perceptions}. While these methods show potential in structured testing, they face barriers in unsupervised use, such as camera calibration difficulties, lighting variability, and privacy concerns, particularly among older adults~\cite{chen2024heuristic,quinn2019mobile}. Furthermore, most mobile interventions are evaluated in short-term trials with heavy researcher support, such as in-person onboarding or mailing preconfigured devices~\cite{mehra2019supporting,mair2022personalized}. This raises questions about whether older adults can independently sustain engagement when such support is absent.

Another strand of research highlights that effective engagement requires attention to motivation and meaning. Takahashi et al. showed that gamified systems combining points, rankings, and group walking effectively increased both activity and social participation~\cite{takahashi2016mobile}. Brox et al. similarly found that online social features in exergames reduced loneliness and improved community connection~\cite{brox2011exergames}. Beyond gamification, notifications, reminders, and motivational messages are common design strategies to reinforce behavior change~\cite{wang2019supporting}, though poorly designed feedback can overwhelm users or reduce motivation. Activity tracking for older adults' well-being must also capture what feels meaningful to them, not just what is measurable, in order to sustain engagement~\cite{wang2024redefining}.

Finally, digital fitness interventions also intersect with mental health and broader well-being. Pywell et al. explored how mobile apps can serve as cost-effective support for addressing the growing demand for accessible mental health support among older adults~\cite{pywell2020barriers}. Marcu and Huh-Yoo~\cite{marcu2024social} emphasized designing for social connectedness to support mental health through meaningful interactions, while Caldeira et al. argued that technology can support self-care and collaborative care, helping older adults maintain independence while staying socially connected~\cite{caldeira2017senior}.

\section{Conclusions}

This study examined the feasibility and engagement of \toolName{}, a mobile fitness app for older adults. In our app-based intervention with 25~adults aged 65--85, participants valued features that made tracking their physical activity effortless and expressed a preference for automatic tracking. Continuous user feedback informed usability refinements, highlighting both opportunities and challenges in digital health interventions. While smartphone-based exercise apps can provide a scalable alternative to traditional programs, their success depends on addressing usability barriers and aligning with users’ existing habits. Older adults may require clearer guidance, simplified interfaces, and familiar tracking methods to fully engage with such interventions. These insights inform the design of accessible technology for aging populations and future applications supporting active lifestyles.

\backmatter

\bmhead{Supplementary information}

Additional information about the study is currently also available on our website\footnote{\url{https://seniorfitstudy.uta.edu/}, Accessed October 2025.}. In case this website is no longer available, various versions of our website is preserved (like many other websites) in the Internet Archive's Wayback Machine\footnote{\url{https://web.archive.org/web/20250228101454/https://seniorfitstudy.uta.edu/}, Accessed October 2025.}.

\bmhead{Acknowledgements}

See Funding section.

\section*{Declarations}

\subsection{Funding}

The research leading to these results received funding from the W.W. Caruth Jr. Fund at the Communities Foundation of Texas (project title: ``Motivational Technology to Increase Physical Activity'').

\subsection{Conflicts of Interest}

Christoph Csallner has a potential research conflict of interest due to a financial interest with Microsoft and The Trade Desk. A management plan has been created to preserve objectivity in research in accordance with UTA policy.

\subsection{Ethics Approval And Consent to Participate}

This study was conducted in accordance with ethical guidelines for research involving human subjects. All research personnel completed
human subjects protection training, and the study protocol (protocol number 2021-0335) was approved by The University of Texas at Arlington's Institutional Review Board (IRB).

Participants provided informed consent before enrollment, ensuring their voluntary participation and a clear understanding of the study’s purpose and procedures.

\subsection{Consent for Publication}

All authors have given consent for publication.

\subsection{Data Availability}

The survey results and \toolName{} app usage data cannot be shared publicly due to privacy (e.g., the GPS walking data would allow locating the participants' precise living location and thus de-anonymize our participants). Upon reasonable request to the authors in the future, a part of the data will be shared on a case-by-case basis.

\subsection{Materials Availability}

All study materials, including our recruitment flyer as well as screenshots of our recruitment website, application form, and survey questionnaires, are available online on Figshare~\cite{SeniorFit_Materials}.

\subsection{Code Availability}

The code for the \toolName{} app is publicly available both on GitHub\footnote{\url{https://github.com/SonicHedghog/Senior-Fit}} and as a permanent archive on Figshare~\cite{SeniorFit_Materials}.

\balance

\subsection{Author Contribution}

Kate Hyun, Kathy Siepker, Xiangli Gu, Angela Liegey-Dougall, Christoph Csallner, and Stephen Mattingly supervised the entire project and developed the IRB protocol. Troyee Saha, Kimberly Vanhoose, Jobaidul Boni, and Samantha Moss were responsible for participant recruitment, phase~1 material development, and conducting the interviews and surveys. Sabrina Haque and Kyle Henry developed the \toolName{} app, and them, along with Christoph Csallner, formulated the app-related survey questionnaires. Sabrina Haque and Christoph Csallner wrote the main manuscript. All authors 
reviewed the manuscript.

\bibliography{ref}

\end{document}